\documentclass[12pt]{article}
\usepackage{amssymb,amsmath,epsfig,graphicx,cite,color}

\numberwithin{equation}{section}
\numberwithin{table}{section}

\def \BB {{\widehat{\cal B}}}

\begin{document}

\baselineskip=16pt

\begin{titlepage}
{}~ \hfill\vbox{\hbox{MAD-TH-07-06} }\break
\vskip 4.1cm

\centerline{\Large \bf Rolling to the tachyon vacuum in string field theory} \vspace*{2.0ex}
\vspace*{8.0ex}

\centerline{\large \rm Ian Ellwood}

\vspace*{8.0ex}

\centerline{\large \it Department of Physics, } 
\centerline{\large \it University of Wisconsin, Madison, WI 53706, USA} 
\vspace*{2.0ex}
\centerline{E-mail: {\tt iellwood@physics.wisc.edu}}

\vspace*{6.0ex}

\vspace*{6.0ex}

\centerline{\bf Abstract}
\bigskip

\noindent
We argue that the rolling-tachyon solution in cubic OSFT proceeds at
late times to precisely the analytic tachyon-vacuum solution constructed
by Schnabl.  In addition, we demonstrate the relationship between the
rolling-tachyon solution and the standard BCFT description by showing that
there is a finite gauge transformation which relates the two.

\end{titlepage}
\tableofcontents


\section{Introduction}
\label{s:intro}

Recently, there has been considerable progress in understanding the
vacuum structure of Witten's cubic string field theory
\cite{Witten:1985cc} following Schnabl's construction of an exact
solution of the equations of motion representing the open-string
tachyon vacuum \cite{Schnabl:2005gv}.  Using this solution, it is
possible to show that the tachyon vacuum has the correct energy
\cite{Schnabl:2005gv,Okawa:2006vm,Fuchs:2006hw} and the expected lack
of physical states \cite{Ellwood:2006ba}\footnote{It is worth pointing
out that the analytic proof of vanishing cohomology of the BRST
operator in \cite{Ellwood:2006ba} has yet to be reconciled with the
numerical evidence (in a different gauge) of states in the cohomology
at non-standard ghostnumber \cite{Imbimbo:2006tz}.}, proving
analytically what had only been known from numerical studies
\cite{Kostelecky:1989nt,Sen:1999nx,Moeller:2000xv,Gaiotto:2002wy,Ellwood:2001py,Ellwood:2001ig,Giusto:2003wc,Imbimbo:2006tz}.

Adding to this OSFT revival is the construction of an exact solution
representing the dynamical rolling of the tachyon from the
perturbative vacuum to the tachyon vacuum
\cite{Schnabl:2007az,Kiermaier:2007ba}.  Although we will focus on the bosonic case, a rolling-tachyon solution has also been constructed
for Berkovits' supersymmetric open string field theory \cite{Berkovits:1995ab,Berkovits:1998bt,Berkovits:2001nr} in
\cite{Okawa:2007ri,Erler:2007rh,Okawa:2007it}.  See also \cite{Fuchs:2007yy} for another approach to marginal 
deformations.

Rolling-tachyon solutions in string field theory have long been
somewhat mysterious.  Numerical attempts to construct such solutions
in OSFT using Feynman-Siegel gauge \cite{Coletti:2005zj}, as well as in $p$-adic
string theory \cite{Moeller:2002vx,Moeller:2003gg}, and in vacuum
string field theory \cite{Fujita:2003ex,Fujita:2004ha} give the
unexpected result that the tachyon does not roll to its value at the
tachyon vacuum, but instead begins to oscillate wildly. 
Perhaps not surprisingly, a similar story holds for the new analytic
solutions, as shown in \cite{Schnabl:2007az,Kiermaier:2007ba}.  While
it is true that even for the exact solutions the computation of the tachyon coefficient is only
numerical, it seems unlikely that an analytic result would eliminate this
unwanted behavior.

We thus have a puzzle: How do we reconcile the strange behavior of the
rolling-tachyon solution with our intuition that the rolling tachyon
should take us from the perturbative vacuum to the tachyon
vacuum?

One answer to this puzzle is that, although the OSFT solutions do
limit to the tachyon vacuum, the wild oscillations are not physical, but due to a  complicated time-dependent gauge transformation.  Indeed, in
\cite{Coletti:2005zj} it was argued that, using such a 
gauge transformation, one can reduce the time-dependence of the
tachyon to simply $e^{X^0}$, reproducing the boundary conformal field theory (BCFT) description \cite{Sen:2002nu,Strominger:2002pc,Larsen:2002wc,Constable:2003rc}.  As
one of the simple results of this paper, we will prove this result
analytically, showing that the rolling solutions are, in some sense, no
more or less complicated that the BCFT deformation.

This resolution of the puzzle, however, is not particularly
satisfying.  One of the beautiful features of OSFT is that the tachyon
vacuum is not a singular field configuration at the boundary of field
space as it is in BCFT.  It is this finiteness that allows one, for
example, to have control over the spectrum of states at the tachyon
vacuum, something which is relatively difficult to see in the BCFT
perspective.

This resolution is also somewhat at odds with the fact that both the
rolling solution and the tachyon vacuum are {\em in the same
gauge}. It is true that the relevant gauge, $\mathcal{B}_0$-gauge, is
not a perfect gauge\footnote{Indeed, one can check that, around the perturbative vacuum, there is one exact state in $\mathcal{L}_0$ level truncation which preserves the gauge; $\mathcal{B}_0 Q_B (\mathcal{L}_0^{\vphantom{\star}}+\mathcal{L}_0^\star) c_1 |0\rangle = 0$ \cite{Schnabl:2005gv}.  Finding a good gauge in OSFT seems to be a difficult problem.  There is also numerical evidence that even Feynman-Siegel gauge is not a good gauge globally \cite{Ellwood:2001ne}.}, but, nonetheless, it greatly restricts the possible gauge transformations.  This suggests
another resolution to the puzzle: the rolling-tachyon solution {\em
does} limit to the tachyon vacuum in spite of all the the numerical
evidence to the contrary\footnote{A third possibility is, of course,
that the rolling solution does not limit to the tachyon vacuum at all, even up to
a gauge transformation, but we will not consider this possibility.}.

It is the main objective of this paper to give evidence for this
resolution.  Indeed we will show how one can find the Schnabl
solution by taking the $X^0 \to \infty$ limit of the rolling solution using some
simplifying assumptions.
Our derivation will be subject to two caveats:
\begin{enumerate}
  \item Unlike in the numerical computations of the tachyon vev, we
will will work in the coordinate system $z = f(w) = \frac{2}{\pi}\arctan(w)$.  We will,
thus, think of quantities as being expanded in a basis of
$\mathcal{L}_0 = f^{-1} \circ L_0$ eigenstates rather than $L_0$
eigenstates.  The transformation between these two descriptions is
quite non-trivial and introduces many potential divergences.  We
suspect that these may play a role in explaining the apparent
inconsistency between our results and the numerical results.
  \item An exact computation of the time-dependence of the rolling
solution in $\mathcal{L}_0$-basis does not appear to be much easier
than in $L_0$-basis.  As such, we make an assumption about the
late-time behavior of the matter correlators, which simplifies the
computation enough that we can find analytic expressions.  This assumption is
specified in (\ref{eq:assumption}).  We consider the fact that using this 
simple assumption leads to Schnabl's solution as a hint that it is probably true.
\end{enumerate}

Having argued that the late-time limit is just the tachyon vacuum, the reader may wonder how the energy of the original brane could possibly be conserved.  Indeed, in a standard classical system, this would be impossible for the following reason:  Suppose we have a time-dependent configuration which at late-times limits to a static configuration.  Since, at late-times, the time-dependent solution becomes approximately constant, the kinetic energy must go to zero.  Hence all of the energy will come from the potential energy, which should be the same as for the static solution.

OSFT violates two assumptions in this argument.  First, as OSFT has an infinite number of time-derivatives, it is possible for the kinetic energy to remain finite even as the solution becomes constant.  Second, the potential of OSFT is not smooth.  In the argument above, we assumed that if two configurations were very close to each other, they would have the same potential energy.  However, in OSFT, we can find solutions which are arbitrarily close to each other in the Fock-space expansion yet have different energies, as is demonstrated by the remarkable fact that the tachyon-vacuum solution is actually a limit of pure-gauge solutions \cite{Schnabl:2005gv, Okawa:2006vm}.

This pathology is related to the lack of a proper norm on the free-string Fock-space that we are using for our classical field space.  Without such a norm, we cannot give a rigorous definition of when two states are close to each other.  The best we can do is see if the coefficients of two states in the level-expansion are near each other.  This definition is not independent of which basis we use, however, and any statement we are making about the late-time limit of the rolling tachyon should be understood to be subject to this important subtlety.

The organization of this paper is as follows: In
section~\ref{s:review}, we review Schnabl's exact expression for the
tachyon vacuum and the rolling-tachyon solution.  Then, in section~\ref{s:limit}, we argue that the late-time limit of the rolling-tachyon solution is given by the tachyon-vacuum solution. Finally, in section~\ref{s:BCFT}, we show how the rolling-tachyon solution is related to the BCFT
deformation, $J = e^{X^0}$.

\section{The tachyon-vacuum and rolling-tachyon solutions}
\label{s:review}

\begin{figure} 
\centerline{
\begin{picture}(299,131)(0,-15)
\includegraphics{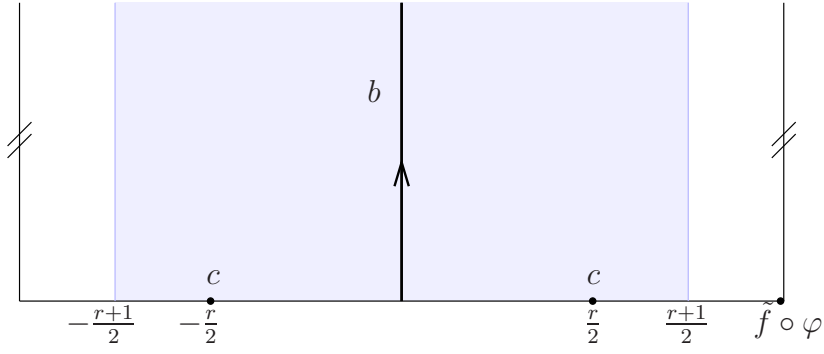}
\end{picture}
\begin{picture}(0,0)(299,-15)
\put(72,8){$c$}
\put(61,-10){$-\frac{r}{2}$}
\put(19,-10){$-\frac{r+1}{2}$}
\put(216,8){$c$}
\put(215,-10){$\frac{r}{2}$}
\put(245,-10){$\frac{r+1}{2}$}
\put(133,77){$b$}
\put(279,-10){$\tilde{f}\circ \varphi$}
\end{picture}
}
\caption{\label{f:tachyonVacuum} The geometric definition of the states $|\psi_n\rangle$.}
\end{figure}

We begin with a short review of the tachyon-vacuum and rolling-tachyon
solutions\footnote{We warn the reader that there are a number of different conventions for defining states in the cylinder coordinate system.  We follow the convention in which the left half of an operator acts as $\mathcal{O}^L (\psi_1 *\psi_2) = (\mathcal{O}^L\psi_1) *\psi_2$.  However, when we display our states graphically, as in figure \ref{f:tachyonVacuum}, the left half of the string is on the right half of the shaded region.  We are also including an extra factor of $\frac{2}{\pi}$ in our conformal map \cite{Okawa:2006vm,Okawa:2006sn,Kiermaier:2007ba}, which is why we do not have the factors of $\pi$ present in the diagrams of \cite{Schnabl:2005gv}.  When we refer to operators such as  $\mathcal{L}_0$ and $\mathcal{B}_0$,  we define them as pull-backs of the non-curly versions: $\mathcal{L}_0 = f^{-1}\circ L_0$.  This definition coincides with the one  in \cite{Schnabl:2005gv}, since the extra numerical factors cancel.}.  Readers unfamiliar with this material should consult \cite{Schnabl:2005gv,Schnabl:2007az,Kiermaier:2007ba}.  
It is convenient to define string field theory states not on the upper
half plane, as is standard in ordinary CFT, but, instead, on the
semi-infinite cylinder $C_r$, which is defined as follows: one takes the
region of the UHP $- r/2 \le \Re(z) \le r/2$ and glues the line
$\Re(z) = - r/2$ to the line $\Re(z) = r/2$.  To define correlation
functions on $C_\alpha$, one uses that
\begin{equation}
   z = f_r(w) =\frac{r}{\pi} \arctan(w)
\end{equation}
maps the UHP to the cylinder $C_r$.  For convenience, we define $f(w)
= f_2(w) = \frac{2}{\pi} \arctan(z)$.

\begin{figure} 
\centerline{
\begin{picture}(367,130)(0,-15)
\includegraphics{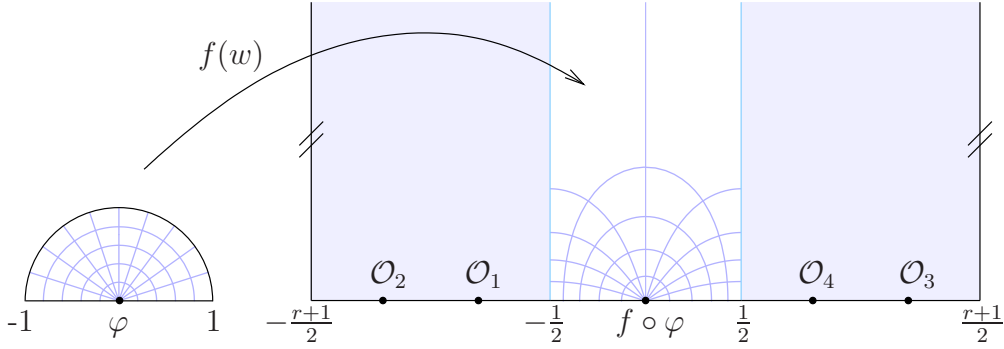}
\end{picture}
\begin{picture}(0,0)(367,-15)
\put(28,-10){$\varphi$}
\put(220,-10){$f\circ \varphi$}
\put(265,-10){$\frac{1}{2}$}
\put(350,-10){$\frac{r+1}{2}$}
\put(185,-10){$-\frac{1}{2}$}
\put(87,-10){$-\frac{r+1}{2}$}
\put(65,-10){1}
\put(-10,-10){-1}
\put(127,8){$\mathcal{O}_2$}
\put(163,8){$\mathcal{O}_1$}
\put(289,8){$\mathcal{O}_4$}
\put(327,8){$\mathcal{O}_3$}
\put(62,91){$f(w)$}
\end{picture}
}
\caption{\label{f:arctanCoords}Here we illustrate how we can define a state $|\chi\rangle$
in the cylinder coordinates.  We begin by mapping the state
$|\varphi\rangle$ into the cylinder geometry using $f(w) =
\frac{2}{\pi} \arctan(w)$.  We then insert the some local operators,
$\mathcal{O}_i$, and compute the correlator on the cylinder.  The
resulting amplitude is defined to be $\langle \varphi |\chi\rangle$
for some state $|\chi\rangle$.}
\end{figure}

We can define states in this coordinate system through their inner
products with arbitrary states, $\varphi$.  For example, we might
define a state $\chi$ through
\begin{equation} \label{eq:wedgeStateWithInsertions}
   \langle \varphi |\chi \rangle   = \langle f \circ \varphi(0) \, \mathcal{O}_1(z_1)  \ldots \mathcal{O}_n(z_n) \rangle_{C_{r+1}} \ ,
\end{equation}
where the $\mathcal{O}_i$ are a set of local operators inserted in
$C_r$.  In order for $\chi$ to be a well-defined state, we should insist
that none of the $z_i$ are in the region $-1/2\le\Re(z)\le1/2$, which
is the image of the unit disk under $f(w)$ and is known as the {\em
coordinate patch}.  A state $|\chi\rangle$ defined through
(\ref{eq:wedgeStateWithInsertions}) is said to be a {\em wedge state
(of width $r$) with insertions} \cite{Schnabl:2002gg,Rastelli:2000iu}.  See figure \ref{f:arctanCoords}.

As we defined things in (\ref{eq:wedgeStateWithInsertions}), the
coordinate patch is in the middle of the cylinder. Since we are more
interested in the part of $C_{r+1}$ that is {\em not} contained in the
coordinate patch (i.e. the shaded region in figure
\ref{f:arctanCoords}), we will rotate the cylinder, $z \to z+
\frac{r+1}{2}$, so that half of the coordinate patch is on right side
of $C_{r+1}$ and half is on the left, while the shaded region is in the
middle.  We denote the map of $\varphi$ into the translated coordinate
patch by $\tilde{f}$.

In addition to inserting local operators on the cylinder, we also need
to insert contour integrals of operators.  In particular, we will
use\footnote{This operator is denoted $B_1^L$ in \cite{Schnabl:2005gv}.}
\begin{equation}
   B= \int_{\gamma} dz \, b(z) \ ,
\end{equation}
where $\gamma$ is the contour $\Re(z) = \text{constant}$, and the direction of
integration is upward.  Since the contour can be freely pushed to the
left or right unless it crosses some other operator, we need only to specify that the contour lies
between the neighboring operators in a given expression.

To define the tachyon vacuum, we define the states $|\psi_n\rangle$ by
\begin{equation}
  \langle \varphi | \psi_n \rangle = \left\langle \tilde{f} \circ \varphi(0) \,\, c(\tfrac{n}{2}) \, B \, c(-\tfrac{n}{2}) \right\rangle_{C_{n+2}}.
\end{equation}
This state is pictured in figure \ref{f:tachyonVacuum}.  The tachyon vacuum is given
by
\begin{equation} \label{eq:SchnablSolution}
  \Psi = \lim_{N \to \infty} \left( \psi_N - \sum_{n = 0}^N \partial_n
  \psi_n \right) \ .
\end{equation}

\begin{figure} 
\centerline{
\begin{picture}(371,131)(0,-15)
\includegraphics{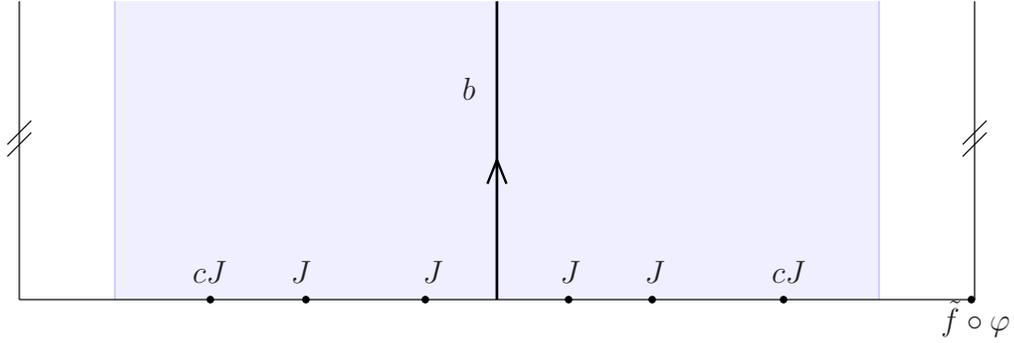}
\end{picture}
\begin{picture}(0,0)(371,-15)
\put(67,8){$cJ$}
\put(104,8){$J$}
\put(154,8){$J$}
\put(206,8){$J$}
\put(238,8){$J$}
\put(286,8){$cJ$}
\put(169,77){$b$}
\put(350,-10){$\tilde{f}\circ \varphi$}
\end{picture}
}
\caption{\label{f:RollingTachyon} The geometric definition of the
states $|\theta_n\rangle$.  The distance between the $J$'s is
integrated from 0 to 1.  For convenience, we have also used the fact
that $B^2 = 0$ to reduce the number of $b$ contours to just one, while
removing all but two of the $c$'s.}
\end{figure}

The rolling solution is a bit more complicated to define in this
notation, although geometrically it is just as elegant.  We start with
our weight one primary $J = e^{X^0}$.  We then define the variables,
\begin{equation}
  t_i =   \tfrac{1}{2} \sum_{j = 1}^{i-1} w_j - \tfrac{1}{2} \sum_{j = i}^{n-1} w_j  \ ,
  \qquad
 r(w_i) = 2+ \sum_{i = 1}^{n-1} w_i  \ ,
\end{equation}
and the states $|\theta_n\rangle$ by
\begin{equation} \label{eq:thetan}
  \langle \varphi| \theta_n \rangle 
   = (-1)^{n+1}\int_0^1 \biggl(\,\prod_{i = 1}^{n-1} d w_i \biggr)
   \left \langle\tilde{f}\circ \varphi(0)\,\,
     cJ(t_n) \, B\,  cJ(t_{n-1})\,  B \ldots B\, cJ(t_1)  
   \right\rangle \ .
\end{equation}
These states are picture in figure \ref{f:RollingTachyon}.

The marginal solution is then given by
\begin{equation}\label{eq:rollingTachyonSolution}
  \Theta = \sum_{n = 1}^\infty  \lambda^n \theta_n \ .
\end{equation}
As is easy to check, the marginal parameter $\lambda$ can be rescaled
by a translation of $X^0$.  The only thing one cannot change in this
way is the {\em sign} of $\lambda$ which must be positive for the
solution to roll towards the tachyon vacuum.  From now on we will
simply set $\lambda = 1$.

\section{The late-time limit of the rolling-tachyon solution}
\label{s:limit}

Having defined the relevant fields, we now argue that, at late times,
the rolling-tachyon solution limits to tachyon vacuum.  As is evident
from the expression for $\Theta$ given in (\ref{eq:thetan}) and
(\ref{eq:rollingTachyonSolution}), a direct attempt to take the limit
$X^0 \to \infty$ would be very difficult.  Indeed, it is not even
obvious that such a limit exists.

However, as we will now show, one finds very nice results if one {\em
assumes} that a limit exists.  In detail, suppose we take the all of
the contributions from $\Theta$ that have a width $r+1$ and sum them
up to give a state $W_r$.  For such a state, the ghost insertions are
fixed and one integrates over various possible insertions of
$e^{X^0(\sigma)}$.  Summing up all the possibilities yields some (very
complicated) functional $\mathcal{F}_r[X^0(\sigma)]$ and we can write
\begin{equation} \label{eq:Wr}
  \langle \varphi | W_r \rangle  = \left \langle \tilde f\circ \varphi \,\,
    c(r/2) \,B \,\mathcal{F}_r[X^0(\sigma)]\, c(-r/2)
  \right \rangle_{C_{r+2}} \ .
\end{equation}
We then make the following
\begin{equation} \label{eq:assumption}
  \text{\bf assumption:}  \qquad \lim_{x^0 \to \infty} \mathcal{F}_r[X^0(\sigma) + x^0] = f(r)  \ ,
\end{equation}
where $f(r)$ is some yet to be determined function.  Note that this
assumption is stronger than the assumption that there exists
a limit.  We are also assuming that the limit does not depend on
operators like $\partial X^0(\sigma)$.  The power of this assumption
is that it implies that if we are only interested in late-time
questions, we can replace all of the explicit $X^0(\sigma)$'s by the
zero mode $x^0$, which is just a constant and not a field.

Replacing $X^0(\sigma) \to x^0$ in (\ref{eq:Wr}) gives
\begin{equation}
  \langle \varphi | W_r \rangle  = \mathcal{F}_r[x^0(\sigma)] \left \langle \tilde f\circ \varphi \,\,
    c(r/2) \,B \, c(-r/2)
  \right \rangle_{C_{r+2}}  \ ,
\end{equation}
which reveals that
\begin{equation}
  |W_r\rangle = \mathcal{F}_r[x^0(\sigma)] \psi_r \ .
\end{equation}
Now $\mathcal{F}_r[x^0]$ is given by the sum over $n$ of the integral over all possible
ways of dividing an interval of width $r$ into $n$ intervals with
width $\le 1$ multiplied by $(-1)^{n}e^{(n+1) X^0}$.  Explicitly,
\begin{equation}
  \mathcal{F}_r[x^0] = \sum_{n = 0}^\infty  (-1)^{n} e^{(n+1)x^0}\biggl(\,\prod_{j = 1}^n \int_0^1 dw_j\biggr) \delta(\sum w_j - r)  \ .
\end{equation}
To evaluate this sum, we Fourier-transform the delta-function,
\begin{multline}
  \mathcal{F}_r[x^0] = 
    \frac{1}{2\pi} \int_{-\infty}^\infty dy\,\sum_{n = 0}^\infty (-1)^n e^{(n+1)x^0}\biggl(\,\prod_{j = 1}^n \int_0^1 dw_j\biggr) \exp(i y (\sum w_j - r))  
    \\
     =    \frac{1}{2\pi} \int_{-\infty}^\infty dy\,\sum_{n = 0}^\infty(-1)^n e^{(n+1)x^0} e^{-i r y}  \left( \frac{1}{iy} (e^{i y} - 1)\right)^n \ .
\end{multline}
Performing the sum over $n$ yields
\begin{equation}
  \mathcal{F}_r[x^0] = \frac{1}{2\pi} \int_{-\infty}^\infty dy\,   \frac{e^{x^0} e^{-i ry}}{1+\frac{1}{iy} (e^{i y} - 1)e^{x^0}} \ .
\end{equation}
We can now take the large $x^0$ limit to find
\begin{equation}
f(r) = \lim_{x^0 \to \infty}  \mathcal{F}_r[x^0]  = \int_{-\infty}^\infty dy\, \frac{(-i y) \,e^{-i  y r}}{1- e^{iy }}
 = \partial_r \int_{-\infty}^\infty dy\, \frac{ \,e^{-i  y r}}{1- e^{iy }} \ ,
\end{equation}
which reduces to
\begin{equation}
  f(r) = \sum_{n = 0}^{\infty}  \delta'(r - n) \ .
\end{equation}
Since, by definition,
\begin{equation}
  \lim_{x^0 \to \infty} \Theta\Bigr|_{X^0 = x^0} = \lim_{x^0 \to \infty} \int_0^\infty dr \,W_r\Bigr|_{X^0 = x^0}  = \int_0^\infty f(r) \psi_r  \ ,
\end{equation}
we learn that
\begin{equation}
  \lim_{x^0 \to \infty} \Theta\Bigr|_{X^0 = x^0}  = \int_0^\infty dr\, \sum_{n = 0}^\infty \delta'(r-n) \psi_r \  ,
\end{equation}
so that
\begin{equation}
  \lim_{x^0 \to \infty} \Theta\Bigr|_{X^0 = x^0}  =  -\sum_{n = 0}^\infty \partial_n \psi_n = \Psi \ ,
\end{equation}
reproducing the tachyon-vacuum solution.  Although this gives a formal
proof that the tachyon vacuum appears in the limit, the reader may
wonder whether the extra piece $\psi_N$ in (\ref{eq:SchnablSolution})
is being correctly accounted for.  To assure the reader, we note that
we can also perform the limit directly in $\mathcal{L}_0$-level
expansion.  One can verify, for example, that, after replacing $X^0
\to x^0$, the rolling-tachyon solution takes the form, 
\begin{equation}
   \Theta\Bigr|_{X^0 = x^0} =\frac{e^{x^0}}{1+\frac{\pi}{2}e^{x^0}} c_1|0\rangle  + \text{higher $\mathcal{L}_0$-level} \ .
\end{equation}
Taking $x^0 \to \infty$ gives $\frac{2}{\pi} c_1 |0\rangle$  for the lowest level term, reproducing the result of \cite{Schnabl:2005gv}.

As a final note, we would like to address the following concern, which
might make the reader believe that this result is actually trivial: Since the rolling-tachyon solution is in
$\mathcal{B}_0$-gauge and reducing $X^0$ to its zero mode preserves
this condition, it might seem that finding the tachyon vacuum is inevitable, as
there is only one such universal solution.  The problem with this
argument is that, after we replace $X^0$ by its zero mode, we {\em no
longer have a solution to the equations of motion}. It is quite
remarkable if our assumption (\ref{eq:assumption}) is wrong that
taking the limit $x^0 \to \infty$ would yield both a finite state and
a classical solution.

\section{The rolling tachyon and BCFT}
\label{s:BCFT}

Having argued that the tachyon-vacuum solution arises as a limit of the
rolling-tachyon solution, we would now like to point out the simple
relationship between the rolling-tachyon solution in OSFT and the
boundary deformation $J = e^{X^0}$ in BCFT\footnote{For a general theory relating
boundary deformations to SFT solutions see \cite{Sen:1990hh,Sen:1990na,Sen:1992pw,Sen:1993mh,Sen:1993kb}.  See also \cite{Recknagel:1998ih} for a general discussion of boundary deformations.}.  The use of identity states and their relation to deformations of the boundary CFT is similar to \cite{Kluson:2003xu}.

Recall that in boundary conformal field theory, one can deform the
boundary conditions of the theory by a true marginal operator
$\mathcal{V}$ by adding a boundary term to the worldsheet action,
\begin{equation}
  S(X,b,c) \to S + \int d \sigma \mathcal{V}(\sigma) \ ,
\end{equation}
where the integral is performed along the boundary of the world sheet.
This implies that a correlator on the UHP in the deformed theory can
be related to a correlator in the undeformed theory by
\begin{equation} \label{eq:deformingCorrelators}
  \langle \mathcal{O}_1(z_1) \ldots \mathcal{O}_n (z_n) \rangle_{\mathcal{V}}
   =  \langle \mathcal{O}_1(z_1) \ldots \mathcal{O}_n (z_n) e^{\int d\sigma \mathcal{V}(\sigma)}\rangle \ .
\end{equation}
Ordinarily, this is not enough to define the deformed theory since the
right hand side will have various divergences when the $\mathcal{V}$
collide with each other.  Conveniently, for the rolling-tachyon
deformation, $\mathcal{V} = J$, no counterterms are necessary since
\begin{equation}
  J(\sigma_1) J(\sigma_2) = (\sigma_1 - \sigma_2)^2 :J(\sigma_1) J(\sigma_2): \ .
\end{equation}
Let us now compare this BCFT description with the OSFT description.
In OSFT, one does not change the underlying CFT, but, instead shifts
the vacuum $\Psi \to \Psi + \Theta$, where $\Theta$ was given in
(\ref{eq:rollingTachyonSolution}).  If one also constructs the string
field theory around the deformed CFT, which we can call OSFT${}_J$,
then there is some complicated field-redefinition which takes one from
the undeformed theory with a shifted vacuum, OSFT${}_{\Theta}$, to
the theory OSFT${}_J$ in which the CFT is deformed.

What is remarkable about the rolling-tachyon solution is that this
field-redefinition is actually a {\em finite} gauge transformation.  To see
how this works, consider the following string field, $\Theta_0$,
defined through the relation,
\begin{equation} \label{eq:Theta0}
  \langle \varphi | \Theta_0 \rangle = \left\langle \tilde{f}\circ \varphi(0) \,\, cJ(0)\right\rangle_{C_1} \ .
\end{equation}
This is just the identity string field with an insertion of $cJ$ on the boundary\footnote{See \cite{Schnabl:2002gg,Schnabl:2005gv} for the definition of the $U_r$ operators.  We are using ${}^\star$ to denote BPZ conjugation as in \cite{Rastelli:2006ap}.};
\begin{equation} \label{eq:Theta0SchnablNotation}
  \Theta_0 = U_1^\star U_1^{\vphantom{\star}} cJ(0) |0\rangle \ .
\end{equation}
This state satisfies the OSFT equations of motion in a trivial way since
\begin{equation}
  Q_B \Theta_0 = \Theta_0 * \Theta_0 = 0  \ .
\end{equation}
Consider the theory $OSFT_{\Theta_0}$ defined by shifting the vacuum
$\Psi \to \Psi + \Theta_0$.  This theory differs from the old theory
only in a correction to the kinetic term,
\begin{equation}\label{eq:shiftInAction}
  S(\Psi +\Theta_0) = S(\Psi) + \tfrac{1}{2}\int \Psi *[\Theta,\Psi] +
  \text{Constant} \ ,
\end{equation}
which changes the propagator.

\begin{figure} 
\centerline{
\begin{picture}(290,211)(0,0)
\includegraphics{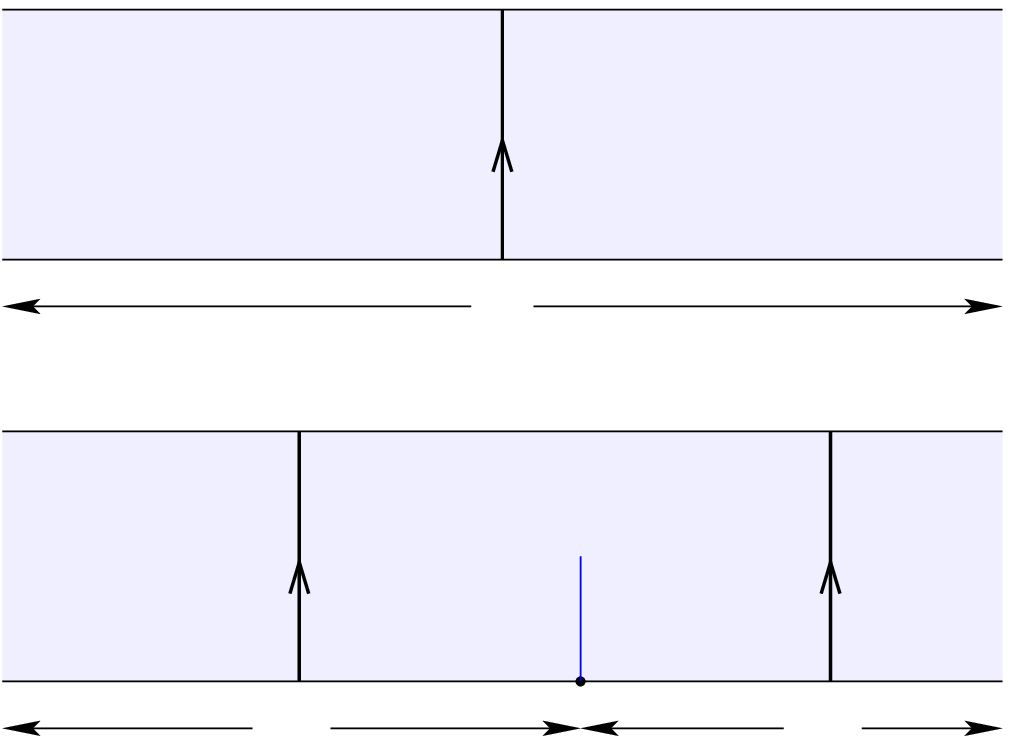}
\end{picture}
\begin{picture}(0,0)(290,0)
\put(168,20){$cJ$}
\put(157,172){$b$}
\put(95,50){$b$}
\put(241,50){$b$}
\put(137,121){$T$}
\put(-40,187){\bf a)}
\put(-40,67){\bf b)}
\put(77,-1){$T_1$}
\put(229,-1){$T_2$}
\end{picture}
}
\caption{\label{f:propagator} In a), the standard Feynman-Siegel gauge
propagator is shown. The modulus $T$ is integrated from zero to
infinity.  In b) the first correction to the propagator from the field
$\Theta_0$ is shown.  Note that there are now two integrals over $b$.
Pulling the right one to the left, one can eliminate the $c$ on the
boundary leaving just $J$.  The two moduli, $T_1$ and
$T_2$ are integrated over which should be thought of as integrating
over the total length of the propagator and the position of the
operator $J$ on the boundary. }
\end{figure}

In Feynman-Siegel gauge, the propagator is just a strip
of worldsheet with one insertion of a line integral of $b$ as shown
in figure \ref{f:propagator}a.  To account for the correction to the propagator
from the modified kinetic term in (\ref{eq:shiftInAction}), we must
include the additional diagrams in which the field $\Theta_0$ is
inserted into the propagator using the cubic vertex.  However, since
$\Theta_0$ is just an identity field with an operator inserted on
its boundary, the modified propagator is just the old propagator with
insertions of $cJ$ on the boundary and a contour
integral of $b(z)$ between each pair of $cJ$'s.  This is illustrated
in figure \ref{f:propagator}b.  By pulling the contour integrals of $b$
to the left we can remove all of the insertions of $c$ (with one
integral of $b$ left over).

After these manipulations, the final propagator is given by the original
propagator with an insertion of $\exp(\int d\sigma J(\sigma))$, which is just the modification of the boundary CFT described
in (\ref{eq:deformingCorrelators}).  It follows that any
correlator in OSFT${}_{\Theta_0}$ is identical to the same correlator
computed in OSFT${}_{J}$, so that the two theories are the same.

What remains to be shown is that the two states, $\Theta$
and $\Theta_0$, are related by a gauge transformation.  We do this by creating a
family of solutions $\Theta_w$ that interpolates between $\Theta_0$ and
$\Theta = \Theta_1$ such that $w$ is a gauge degree of freedom.

The states $\Theta_w$ are simply the reparametrizations of the state
$\Theta$ discussed in \cite{Okawa:2006sn,Erler:2006hw}.  One forms them by the following procedure: If a state
$|\chi\rangle$ is defined by a correlator,
\begin{equation}
  \langle \varphi | \chi \rangle = \langle \tilde{f} \circ \varphi(0) \,\, \mathcal{O}_1(z_1) \ldots \mathcal{O}_n(z_n) \rangle_{C_{r+1}} \ ,
\end{equation}
one can define a new state $\chi_w$ by removing the coordinate patch
from $C_{r+1}$ (leaving a vertical strip of width $r$), shrinking the
remaining vertical strip using $z \to w z$ (so that the strip is now
of width $rw$) and then gluing back in the coordinate patch.  This
yields a correlator on $C_{1+rw}$ which, in turn, defines a state
$|\chi_w\rangle$.

The explicit operator form of this procedure is determined by the identity,
\begin{equation} \label{eq:reparamDef}
  e^{\frac{\beta}{2} (\mathcal{L}_0^{\vphantom{\star}}-\mathcal{L}_0^\star)} \chi_w =\chi_{e^{\beta} w} \ .
\end{equation}
When two states are related by a reparametrization,
they are also related by a gauge transformation.  This immediately
implies that all of the $\chi_n$ for $n>0$ are related by finite gauge
transformations.  However, $\chi_0$ can only be reached by an infinite
reparametrization, taking $\beta \to -\infty$.  Happily, it turns out
that for the rolling-tachyon solution, there is a different gauge
transformation that remains completely finite even as $w \to 0$.

First, however, we should show that $\Theta_w$ at $w = 0$ is the state
$\Theta_0$ that we defined in (\ref{eq:Theta0}).  This is seen by
noting that, as we take $w \to 0$, the regions of integration in the $\theta_n$ (defined in
(\ref{eq:thetan})) shrink to zero size, so that the only term that
survives in this limit is $|\theta_1\rangle$, which is given by
\begin{equation}
    \langle \varphi|\theta_1\rangle = \langle \tilde{f} \circ \varphi \,\, cJ(0)\rangle_{C_2}\ .
\end{equation}
Since the operator $cJ$ is a conformal primary of weight zero, it is
not affected by the rescaling $z \to w z$, which thus has the effect
of reducing $C_2 \to C_1$ as $w \to 0$ so that we recover
(\ref{eq:Theta0}).  Hence we find that the string field $\Theta_0$
introduced in (\ref{eq:Theta0}) is indeed what we get when we use the
reparametrization $\Theta \to \Theta_w$ as $w \to 0$.

We now wish to show that the $\Theta_w$ are all gauge equivalent under
finite gauge transformations, including the case $w = 0$.  We show
this using the following identity, which is straightforward to
prove (see appendix \ref{a:flowIdentity}):
\begin{equation} \label{eq:wedgeAngleFlowEquation}
 -2\, \partial_w \Theta_w = Q_B (\BB \Theta_w) + [\Theta_w , \BB \Theta_w] \ ,
\end{equation}
where $\BB = \mathcal{B}_0^{\vphantom{\star}} + \mathcal{B}_0^\star$ \cite{Schnabl:2002gg,Schnabl:2005gv}.
The right had side should be recognized as an infinitesimal gauge
transformation with gauge parameter $\Lambda = \BB \Theta_w$.  Since
$\BB \Theta_w$ is finite as $w \to 0$,
(\ref{eq:wedgeAngleFlowEquation}) gives a finite gauge transformation
relating $\Theta_0$ to $\Theta_w$ for any $w$.  Indeed, if we want, we
can integrate these infinitesimal gauge transformations using\footnote{Such a path ordered exponential of string fields has also appeared recently in \cite{Okawa:2007it}.}
\begin{equation}
  e^{\Lambda(w)} \equiv P\exp\left(-\frac{1}{2} \int_{0}^w dw'\,  \BB \Theta_{w'}\right) \ ,
\end{equation}
where the $P$ indicates path ordering; when expanding out the
exponential we should always push $\Theta_w$'s with larger $w$ to the
right.  We then have the expression,
\begin{equation}
  \Theta_w = e^{-\Lambda(w)}( \Theta_0 + Q_B )e^{\Lambda(w)} \ ,
\end{equation}
which relates the rolling-tachyon solution to the trivial solution (\ref{eq:Theta0}) by a finite gauge transformation.

\begin{figure} 
\centerline{
\begin{picture}(290,95)(0,-15)
\includegraphics{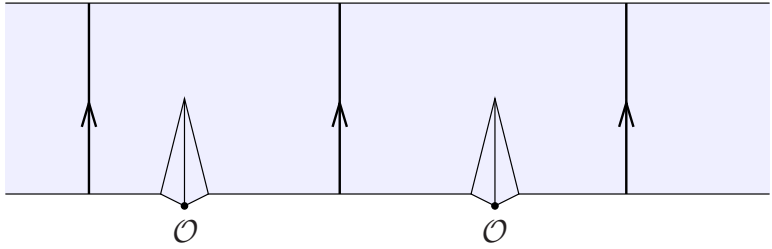}
\end{picture}
\begin{picture}(0,0)(290,-15)
\put(60,-12){$\mathcal{O}$}
\put(177,-12){$\mathcal{O}$}
\end{picture}
}
\caption{\label{f:modifiedPropagator}  The appearance of the propagator with
insertions of a field with small, but finite, width.}
\end{figure}

We close with a few heuristic remarks about the relation
between OSFT and BCFT.  In relating the rolling-tachyon solution to
the BCFT deformation, we used the fact that for the solution
(\ref{eq:Theta0}), the propagator of the theory was modified in
precisely the same way as if we had turned on a boundary deformation.
What happens if we repeat the same argument for the finite-width
states, $\Theta_w$?  Instead of local-operator insertions on the
boundary of the propagator, one inserts pieces of worldsheet as
illustrated in figure \ref{f:modifiedPropagator}.  These extra pieces of worldsheet act as
a cutoff; even when two insertions of $\Theta_w$ collide, the local
operators inside one $\Theta_w$ never get closer than a distance $\sim
w$ to the operators inside another.  This is a very special choice of cutoff that preservers BRST
invariance.  Indeed, it is easy to check that the condition for BRST invariance is just $Q_B \Theta_w + \Theta_w * \Theta_w = 0$, which reproduces the classical equations of motion.

Since $w$ acts as a cutoff on the distance between the local operators
on the boundary, we can think of equations like
(\ref{eq:wedgeAngleFlowEquation}) as being analogous to a
$\beta$-function for the theory since they tell us how the parameters
of the theory flow as we change the scale of the theory.  Moreover, we
can think of the identity limit as being analogous to the {\em
infrared} and the large wedge-angle limit as being the {\em UV}.  In
the deep infrared, the string field reduces to a local operator on the
boundary of the identity and we find a BCFT-like deformation.
Typically, much of the information about the full string field is lost
in this limit so it is not usually possible to reconstruct the full
string field from a knowledge of the BCFT it is associated with by
using an equation like (\ref{eq:wedgeAngleFlowEquation}).  However,
the case of the rolling-tachyon field is special since the operators
involved have a finite OPE.  Because of this, knowing the BCFT
description is enough to reconstruct the full string field by
``flowing to the UV'' using (\ref{eq:wedgeAngleFlowEquation}).

\section*{Acknowledgments}
I would like to thank T.~Grimm and B.~Zwiebach for discussions and M.~Schnabl for useful comments on the draft. This
work was supported in part by DOE grant DE-FG02-95-ER40896 and funds
from the University of Wisconsin-Madison.

\appendix

\section{Proof of the identity (\ref{eq:wedgeAngleFlowEquation})}
\label{a:flowIdentity}
We wish to show
\begin{equation} \label{eq:wedgeAngleFlow}
  -2 \partial_w \Theta_w = Q_B(\BB \Theta_w) + [\Theta_w,\BB] \ .
\end{equation}
We are given the reparametrization identity,
\begin{equation}
  \Theta_{e^{\beta}} = e^{\frac{\beta}{2} (\mathcal{L}_0^{\vphantom{\star}} - \mathcal{L}_0^\star)} \Theta  \ ,
\end{equation}
which yields
\begin{equation} \label{eq:infReparam}
  \partial_w \Theta = \tfrac{1}{2 w} (\mathcal{L}_0^{\vphantom{\star}} - \mathcal{L}_0^\star) \Theta_w \ .
\end{equation}
We are also given the analogue of $\mathcal{B}_0$-gauge for $\Theta_w$:
\begin{equation}\label{eq:B0gaugeForw}
 \left[ \tfrac{1}{2} (\mathcal{B}_0^{\vphantom{\star}} - \mathcal{B}_0^\star) + 
  \tfrac{w}{2} \BB\, \right] \Theta_w = 0 \ .
\end{equation}
Acting on this equation with $Q_B$, we learn that
\begin{equation}
   \tfrac{1}{2} (\mathcal{L}_0^{\vphantom{\star}} - \mathcal{L}_0^\star)  \Theta_w 
   + \tfrac{w}{2} Q_B(\BB  \Theta_w) + \tfrac{1}{2}  (\mathcal{B}_0^{\vphantom{\star}} - \mathcal{B}_0^\star) (\Theta_w * \Theta_w) = 0 \ .
\end{equation}
Using the fact that $(\mathcal{B}_0^{\vphantom{\star}} - \mathcal{B}_0^\star) $ is derivation of the star algebra \cite{Rastelli:2000iu,Schnabl:2005gv}, as well as (\ref{eq:B0gaugeForw}) again, we learn
\begin{equation} \label{eq:finalFlowEquationIdentity}
   \tfrac{1}{2} (\mathcal{L}_0^{\vphantom{\star}} - \mathcal{L}_0^\star)  \Theta_w
    =  - \tfrac{w}{2} Q_B ( \BB \Theta_w) - \tfrac{w}{2} [\Theta_w ,\BB \Theta_w] \ .
\end{equation}
Using (\ref{eq:finalFlowEquationIdentity}) in (\ref{eq:infReparam}) yields (\ref{eq:wedgeAngleFlow}).  It follows that (\ref{eq:wedgeAngleFlow}) holds for all reparametrizations of solutions in $\mathcal{B}_0$-gauge.

\bibliography{c}\bibliographystyle{utphys}

\end{document}